\documentclass[12pt]{article}
\usepackage{epsfig}
\usepackage{graphicx}
\usepackage{color}

\def\Title#1{\begin{center} {\Large #1 } \end{center}}

\newenvironment{Abstract}{\begin{quotation}  }{\end{quotation}}
\newenvironment{Presented}{\begin{quotation} \begin{center} 
             PRESENTED AT\end{center}\bigskip 
      \begin{center}\begin{large}}{\end{large}\end{center} \end{quotation}}

\def\beq{\begin{equation}}
\def\eeq#1{\label{#1}\end{equation}}
\def\eeqn{\end{equation}}

\def\beqa{\begin{eqnarray}}
\def\eeqa#1{\label{#1}\end{eqnarray}}
\def\eeqan{\end{eqnarray}}

\let\bar=\overbar

\def\Dslash{\not{\hbox{\kern-4pt $D$}}}
\def\dslash{\not{\hbox{\kern-2pt $\del$}}}

\def\msb{{\bar{\ssstyle M \kern -1pt S}}}

\newcommand{\pipi}{\ensuremath{B \rightarrow \pi \pi}}
\newcommand{\pippim}{\ensuremath{B^{0} \rightarrow \pi^{+} \pi^{-}}}

\newcommand{\rhorho}{\ensuremath{B \rightarrow \rho \rho}}

\newcommand{\rhozrhoz}{\ensuremath{B^{0} \rightarrow \rho^{0} \rho^{0}}}

\newcommand{\aonepi}{\ensuremath{B^{0} \rightarrow a_{1}(1260)^{\pm} \pi^{\mp}}}

\newcommand{\Ep}{\ensuremath{e^{+}}}
\newcommand{\Em}{\ensuremath{e^{-}}}
\newcommand{\pip}{\ensuremath{\pi^{+}}}
\newcommand{\pim}{\ensuremath{\pi^{-}}}
\newcommand{\aone}{\ensuremath{a_{1}(1260)}}
\newcommand{\Bp}{\ensuremath{B^{+}}}
\newcommand{\Bz}{\ensuremath{B^{0}}}
\newcommand{\Bzb}{\ensuremath{\bar{B}^{0}}}
\newcommand{\YFS}{\ensuremath{\Upsilon(4S)}}

\newcommand{\BBbar}{\ensuremath{B\bar{B}}}

\newcommand{\Dt}{\ensuremath{\Delta t}}

\newcommand{\Acp}{\ensuremath{{A}_{CP}}}

\newcommand{\Scp}{\ensuremath{{S}_{CP}}}

\newcommand{\phione}{\ensuremath{\phi_{1}}}
\newcommand{\phitwo}{\ensuremath{\phi_{2}}}

\newcommand{\rz}{\ensuremath{\rho^{0}}}
\newcommand{\Ks}{\ensuremath{K^{0}_{S}}}
\newcommand{\Kl}{\ensuremath{K^{0}_{L}}}

\begin{document}
\Title{Results for $\phione$ and $\phitwo$ from Belle}
\bigskip\bigskip

\begin{raggedright}  
{\it Pit Vanhoefer\index{Vanhoefer, P.}\\
Max-Plank-Institut f\"ur Physik\\
 F\"oringer Ring 6\\
  M\"unchen 80805 GERMANY}
\bigskip\bigskip
\end{raggedright}


\begin{Abstract}
We present a summary of measurements sensitive to the CKM angles \phione\ and \phitwo, performed by the Belle experiment using the final data sample of $772 \times 10^{6}$ \BBbar\ pairs at the \YFS\ resonance produced at the KEK asymmetric \Ep\Em\ collider. We discuss $CP$ asymmetries from the decays $B^{0}\to c\bar{c} K^{0}$, $D^{(*)+}D^{(*)-}$ which are sensitive to \phione\ and from $B \rightarrow \pi^{+} \pi^{-}$, $a_1{\pm}\pi^{\mp}$ being sensitive to \phitwo. Furthermore the measurement of the branching fraction of $B^{0}\to \rz\rz$ decays and fraction of longitudinal polarization in this decay is presented and used to constrain $\phi_2$ with an isospin analysis in the $B\to \rho\rho$ system.
\end{Abstract}
\vfill
\begin{Presented}
CKM2012, the 7th International Workshop on the CKM Unitarity Triangle\\
Cincinnati, USA, 6-10 September 2012
\end{Presented}
\vfill
\def\thefootnote{\fnsymbol{footnote}}
\setcounter{footnote}{0}

\section{Introduction}
One major precision test of the Standard Model (SM) is to validate the Cabibbo-Kobayashi-Maskawa (CKM) mechanism for violation of the combined charge-parity ($CP$) symmetry~\cite{C,KM}. This is one of the main purposes of the Belle experiment at KEK which has significantly contributed proving the CKM scheme and constraining the unitarity triangle for $B$ decays to its current precision. Any deviation from unitarity would be a clear hint for physics beyond the SM. These proceedings give a summary of the experimental status of measurements of the CKM angles \phione\ and \phitwo\ defined from the CKM matrix elements as $\phione \equiv \pi - \arg(-V_{td}V^{*}_{tb})/(V_{cd}V^{*}_{cb})$ and $\phitwo \equiv \arg(-V_{td}V^{*}_{tb})/(V_{ud}V^{*}_{ub})$.

The CKM angles can be determined by measuring the time-dependent asymmetry between \Bz\ and \Bzb\ decays into a common $CP$ eigenstate~\cite{CP}. In the decay sequence, $\YFS \rightarrow B_{CP}B_{\rm Tag} \rightarrow f_{CP}f_{\rm Tag}$, where one of the $B$ mesons decays into a $CP$ eigenstate  $f_{CP}$ at a time $t_{CP}$ and the other decays into a flavour specific final state $f_{\rm Tag}$ at a time $t_{\rm Tag}$, the time-dependent decay rate is given by
\begin{equation}
  {P}(\Delta t, q) = \frac{e^{-|\Dt|/\tau_{B^0}}}{4\tau_{B^0}} \bigg[ 1+q(\Acp\cos\Delta m_d \Dt + \Scp\sin\Delta m_d \Dt) \bigg],
\label{eq1}
\end{equation}
where $\Dt \equiv t_{CP}- t_{\rm Tag}$, $\Delta m_d$ is the mass difference between the mass eigenstates  $B_{H}$ and $B_{L}$ and $q = +1 (-1)$ for $B_{\rm Tag} = \Bz (\Bzb)$. The $CP$ asymmetry is given by 
\begin{equation}
\frac{N(\bar{B}\to f_{CP}) - N(B\to f_{CP})}{N(\bar{B}\to f_{CP}) + N(B\to f_{CP})},
\label{eq_asym}
\end{equation}
where $ N(B(\bar{B})\to f_{CP})$ is the number of events of a $B(\bar{B})$ decaying to $f_{CP}$, the asymmetry can be time-dependent.
 The parameters \Acp\ and \Scp\ describe direct and mixing-induced $CP$ violation, respectively~\footnote{There exists an alternate notation where $C_{CP} = -\Acp$.}.  All measurements presented here are based on Belle's final data set of $772 \times 10^{6}$ $B\bar{B}$ pairs. 

\section{The Angle \phione}
First-order (tree) weak processes proceeding via $b \rightarrow c$ quark transitions such as $\Bz \to (c\bar{c}) K^0$, $D^{(*)+}D^{(*)-}$, are directly sensitive to the angle \phione, which is the CKM angle currently measured with the smallest experimental uncertainty. 

\subsection{The Decay Channels $B^0\to (c\bar{c})K^0$}
\begin{figure}[htb]
 \begin{minipage}[]{0.45\columnwidth}   
\includegraphics[height=165pt,width=165pt]{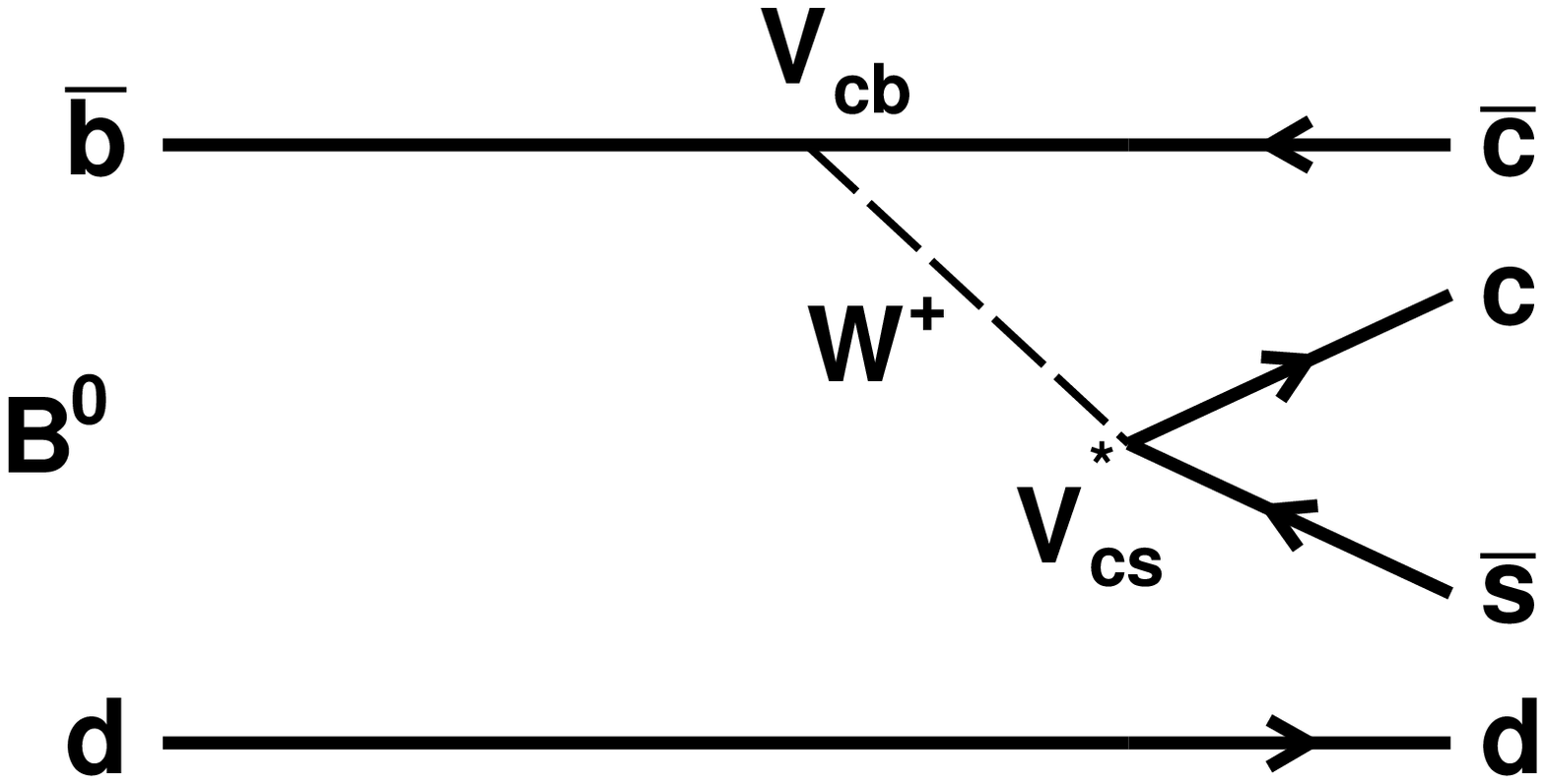}
 \put(-160,20){\scriptsize a)}\\
\end{minipage}
 \begin{minipage}[]{0.53\columnwidth}   
 \includegraphics[height=165pt,width=!]{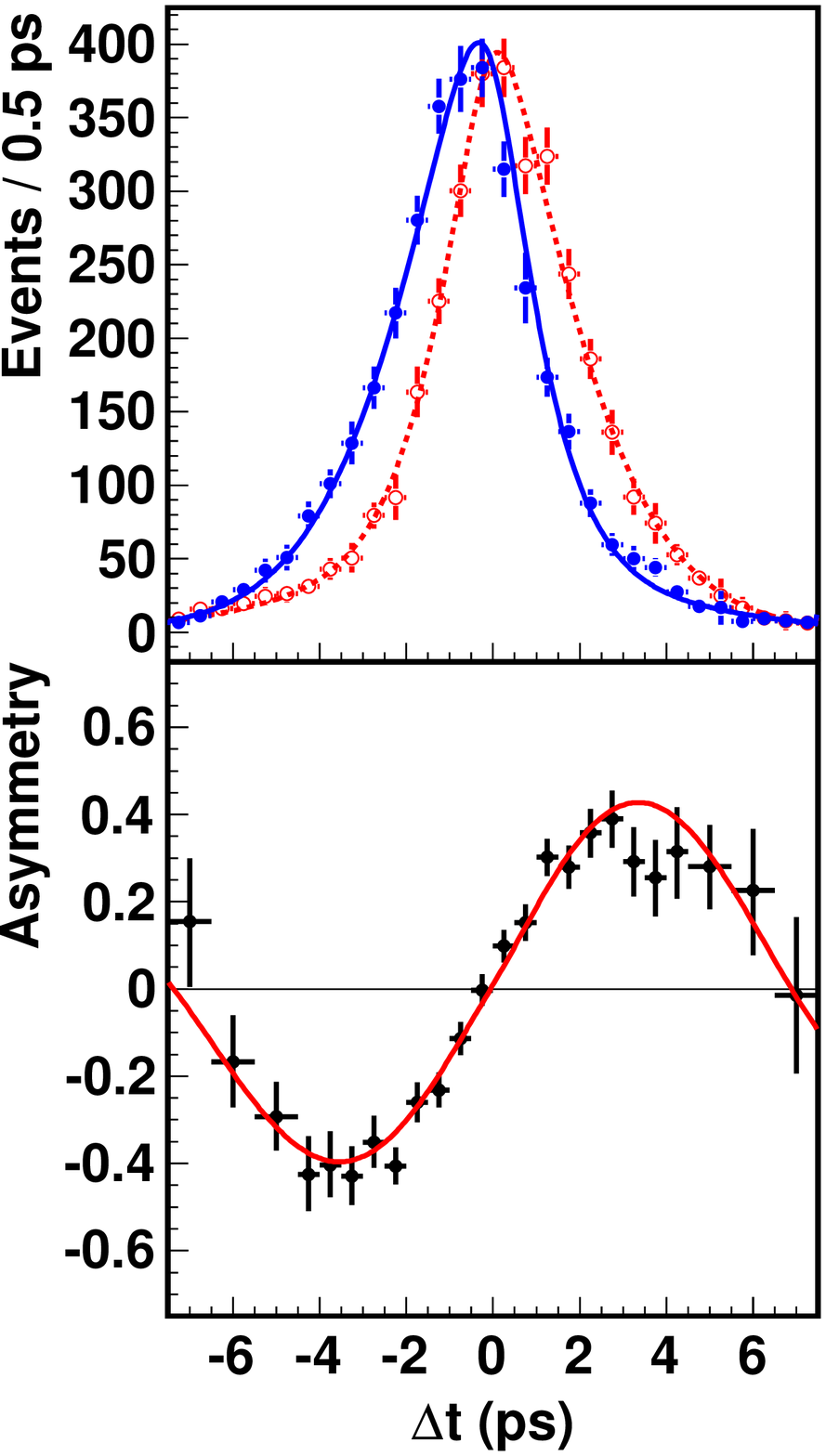}
      \includegraphics[height=165pt,width=!]{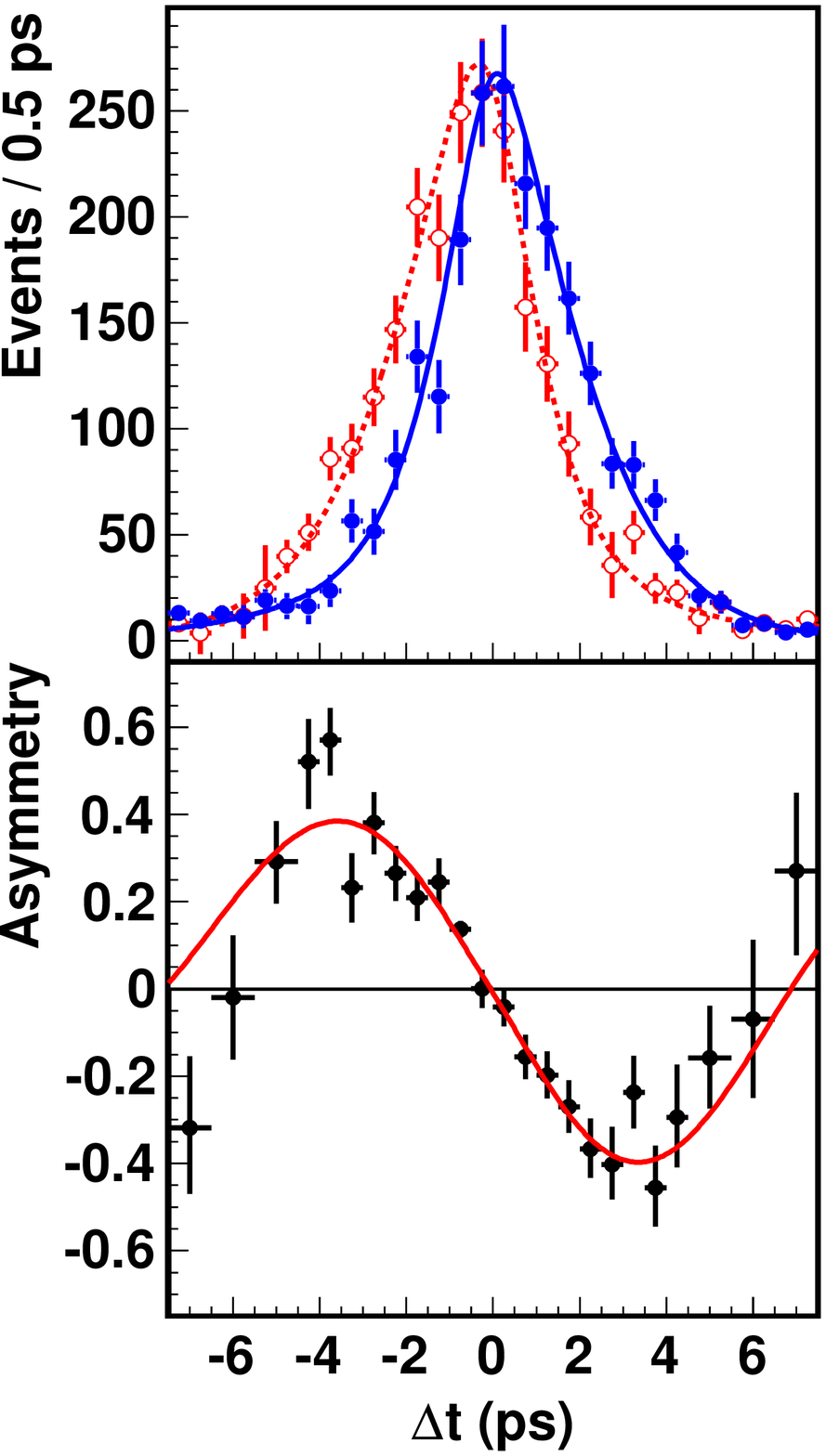}
 \put(-200,100){\scriptsize b)}
\put(-200,20){\scriptsize c)}
\put(-95,100){\scriptsize d)}
\put(-95,20){\scriptsize e)}
\put(-160,150){\includegraphics[height=10pt,width=!]{B-logo.epsf} }
\put(-65,150){\includegraphics[height=10pt,width=!]{B-logo.epsf} }
      \put(-160,80){\textcolor{blue}{\scriptsize$q = -1$}}
      \put(-133,80){\textcolor{red}{\scriptsize$q= +1$}}
      \put(-150,20){\textcolor{black}{\scriptsize$\eta_{CP} = -1$}}
      \put(-55,20){\textcolor{black}{\scriptsize$\eta_{CP} = +1$}}\\
\end{minipage}
 \caption{a) leading-order process of  $B^0\to (c\bar{c})K^0$ decays. b) and d) $\Delta t$ distributions with the fit result on top. c) and e) time-dependent $CP$ asymmetries. b) and c) show the $CP$-odd states and d) and e) show the $CP$-even state $B^{0}\to J/\psi\Kl$. Mixing-induced $CP$ violation can be clearly seen in the asymmetry plots and no height difference in the \Dt\ distributions indicates no direct $CP$ violation.}
  \label{fig_ccK_tcpv}
\end{figure}

The decays $B^0\to (c\bar{c})K^0$, including the so-called `golden channel' $B^0\to J/\psi \Ks$, provide a theoretically and experimentally very clean environment to extract \phione. Since possible additional contributions to the leading order tree diagram, see Fig.~\ref{fig_ccK_tcpv}~a), are negligible or even carry the same weak phase~\cite{JPsiKs_theo}, the measured $CP$ asymmetry ${\cal S}_{CP} = \eta_{CP}\sin(2\phione)$ reveals an unpolluted value of \phione, with the decay channel's $CP$ eigenvalue $\eta_{CP}$. The combined measurement of $CP$ violation in the golden channel and $B^{0} \to \psi(2S)\Ks$, $B^{0} \to\chi_{c}\Ks$, and $B^{0}\to J/\psi\Kl$ provide currently the world's most precise value of $\sin(2\phione) = 0.667 \pm 0.023\;(\rm stat) \pm 0.012\;(\rm syst)$~\cite{ccK_Belle}, as shown in Fig.~\ref{fig_ccK_tcpv}. No direct $CP$ violation was observed, ${\cal A}_{CP} = 0.006 \pm 0.016\;(\rm stat) \pm 0.012\;(\rm syst)$, as predicted by the SM~\cite{BigiSander2}.

\subsection{The Decays Channels $B^0\to D^{(*)+}D^{(*)-}$}
\begin{figure}[htb]
\includegraphics[height=140pt,width=145pt,bb=0 275 416 560,clip=true]{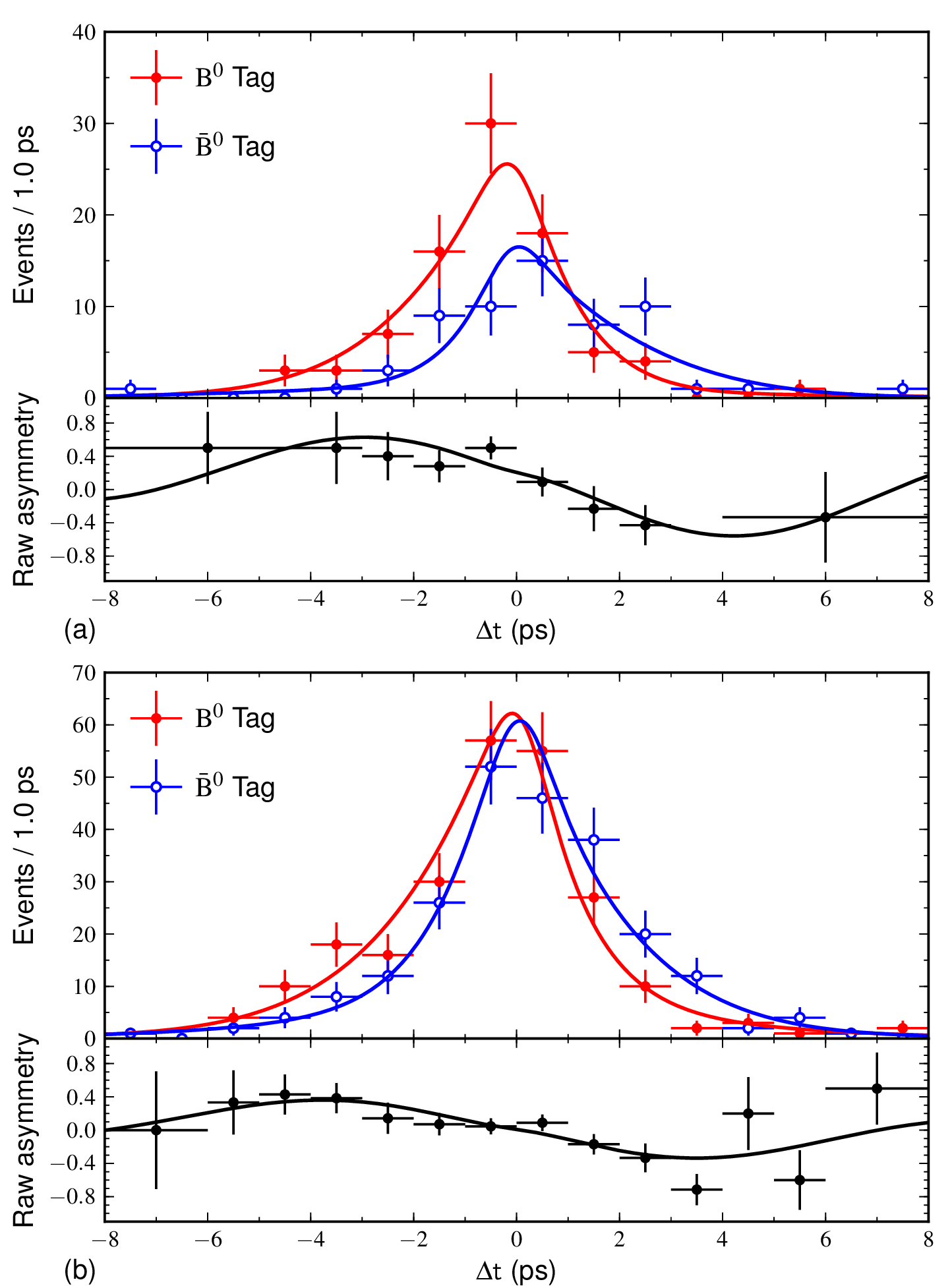}
\includegraphics[height=140pt,width=145pt,bb=0 0 416 275,clip=true]{DpDm_belle.eps}  
\includegraphics[height=140pt,width=!]{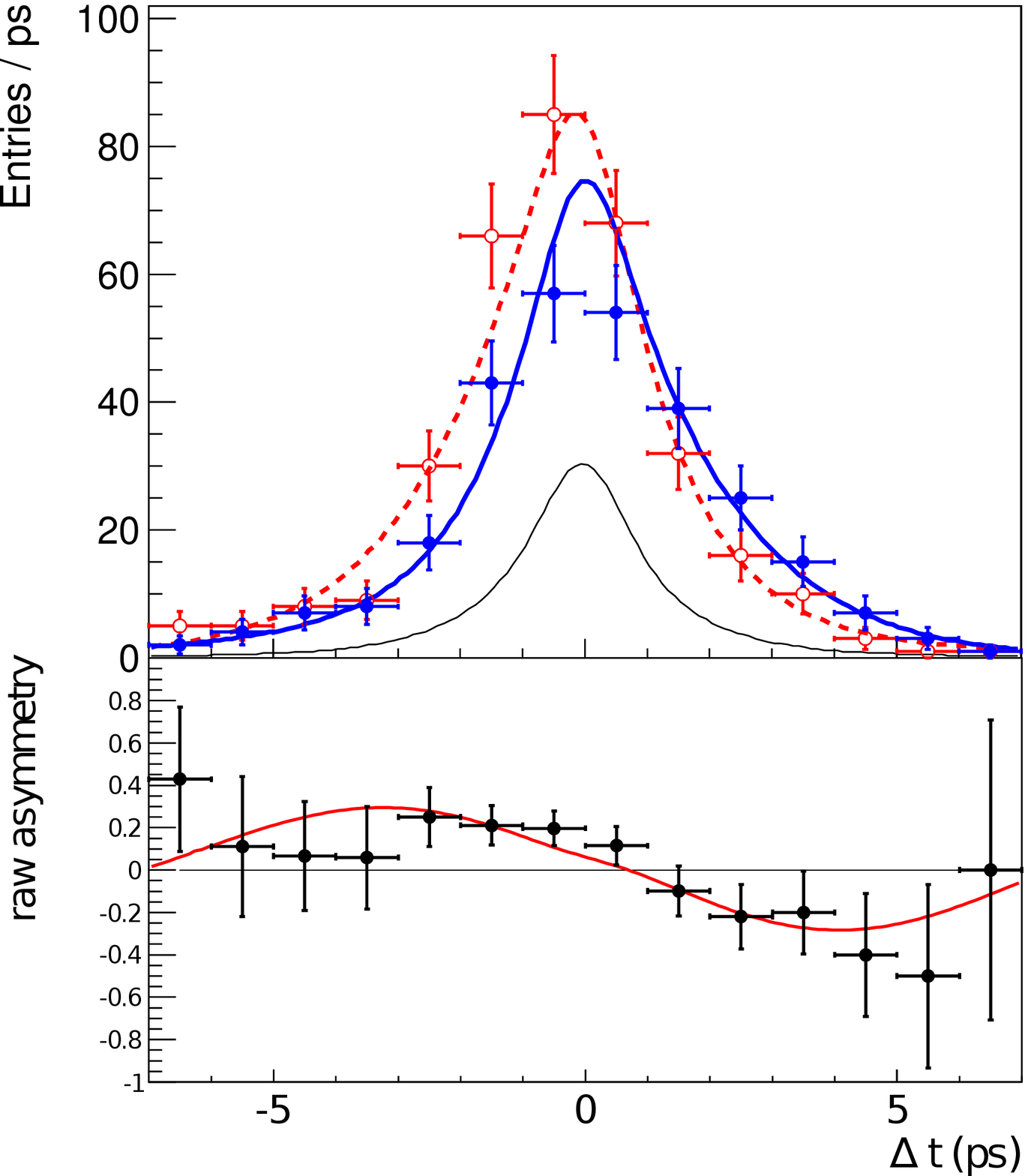}
 \put(-400,1){\scriptsize a)}
 \put(-250,1){\scriptsize b)}
 \put(-110,1){\scriptsize c)}\\
  \caption{ $\Delta t$ distributions for each flavour tag and the $CP$ asymmetries, each with the fit result on top. a) $B^0 \to D^{+}D^{-}$, b) $B^0 \to D^{*\pm}D^{\mp}$ and c) $B^0 \to D^{*+}D^{*-}$.}
  \label{fig_DD_tcpv}
\end{figure}

The decays $B^0\to D^{(*)+}D^{(*)-}$ are also sensitive to $\phione$, however additional contributions from loop (penguin) processes make a pollution of the measured observables possible. Hence also direct $CP$ violation can occur. Compared to the previous Belle measurement, the updated result of $B^0\to D^{+}D^{-}$ decays benefits from a better continuum suppression due to the use of neural-networks. The $CP$ asymmetries obtained are ${\cal S}_{CP} = 1.06^{+0.21}_{-0.14} \;(\rm stat) \pm 0.08 \;(\rm syst)$ and ${\cal A}_{CP} = 0.43 \pm 0.16\;(\rm stat) \pm 0.05\;(\rm syst)$~\cite{DpDm_Belle} and are in good agreement with the results from Babar~\cite{DpDm_Babar}. 
The pseudo-scalar to vector scalar decay $B^0\to D^{*\pm}D^{\mp}$ is a decay into a non $CP$-eigenstate. One therefore has to consider four flavor charge combinations and the time-dependent decay rate in Eq.~\ref{eq1} has to be expanded to five $CP$ parameters~\cite{nonCPeigenstates, nonCPeigenstates2}. The indirect $CP$ asymmetry obtained is $S_{CP} = -0.78 \pm 0.15\;(\rm stat) \pm 0.05\;(\rm syst)$, no direct $CP$ violation has been observed~\cite{DpDm_Belle}.
$B^0\to D^{*+}D^{*-}$ is a pseudo-scalar to vector vector decay and therefore composed of $CP$ even and odd components. An angular analysis in the transversity basis is performed to separate the different $CP$ states. The fraction of $CP$-odd states is found to be $R_{\perp} = 0.138 \pm 0.024\;(\rm stat) \pm 0.006\;(\rm syst)$ and a first observation of mixing-induced $CP$ violation has been reported; $S_{CP} = -0.79 \pm 0.13\;(\rm stat) \pm 0.03\;(\rm syst)$, ${\cal A}_{CP} = 0.15 \pm 0.08\;(\rm stat) \pm 0.04\;(\rm syst)$~\cite{DstrpDstrm_Belle}. 
The $\Delta t$ distributions and $CP$ asymmetries for each mode are shown in Fig.~\ref{fig_DD_tcpv}.

\section{\phitwo}
Decays proceeding via $b \rightarrow u \bar{u} d$ quark transitions such as $\Bz \rightarrow \pi\pi$, $\rho\pi$, $\rho\rho$ and $\aone\pi$, are directly sensitive to \phitwo. At tree level we expect $\Acp=0$ and $\Scp=\sin2\phitwo$. Again, possible penguin contributions can give rise of direct $CP$ violation, $\Acp\neq 0$ and also pollute the measurement of \phitwo, $\Scp=\sqrt{1-\Acp^{2}}\sin(2\phitwo^{eff})$ where the observed $\phitwo^{eff} \equiv \phitwo - \Delta \phitwo$ is shifted by $\Delta \phi_2$ due to different weak phases from additional non-leading contributions.

Despite this, it is possible to determine $\Delta \phitwo$ in $\Bz \rightarrow h^{+} h^{-}$ with an $SU(2)$ isospin analysis by considering the set of three $B \rightarrow hh$ decays where the $hh$s are either two pions or two longitudinally polarized $\rho s$, related via isospin symmetry~\cite{iso}. The $B \rightarrow hh$ amplitudes obey the triangle relations,
\begin{equation}
  A_{+0} = \frac{1}{\sqrt{2}}A_{+-} + A_{00}, \;\;\;\; \bar{A}_{-0} = \frac{1}{\sqrt{2}}\bar{A}_{+-} + \bar{A}_{00}.
  \label{eq_iso}
\end{equation}
Isospin arguments demonstrate that $\Bp \rightarrow h^{+} h^{0}$ is a pure first-order mode in the limit of neglecting electroweak penguins, thus these triangles share the same base, $A_{+0}=\bar{A}_{-0}$. $\Delta \phitwo$ can then be determined from the difference between the two triangles. This method has an inherent 8-fold discrete ambiguity in the determination of \phitwo.

\section{The Decay \pipi}
\begin{figure}[htb]
  \centering
  \includegraphics[width=.49\textwidth]{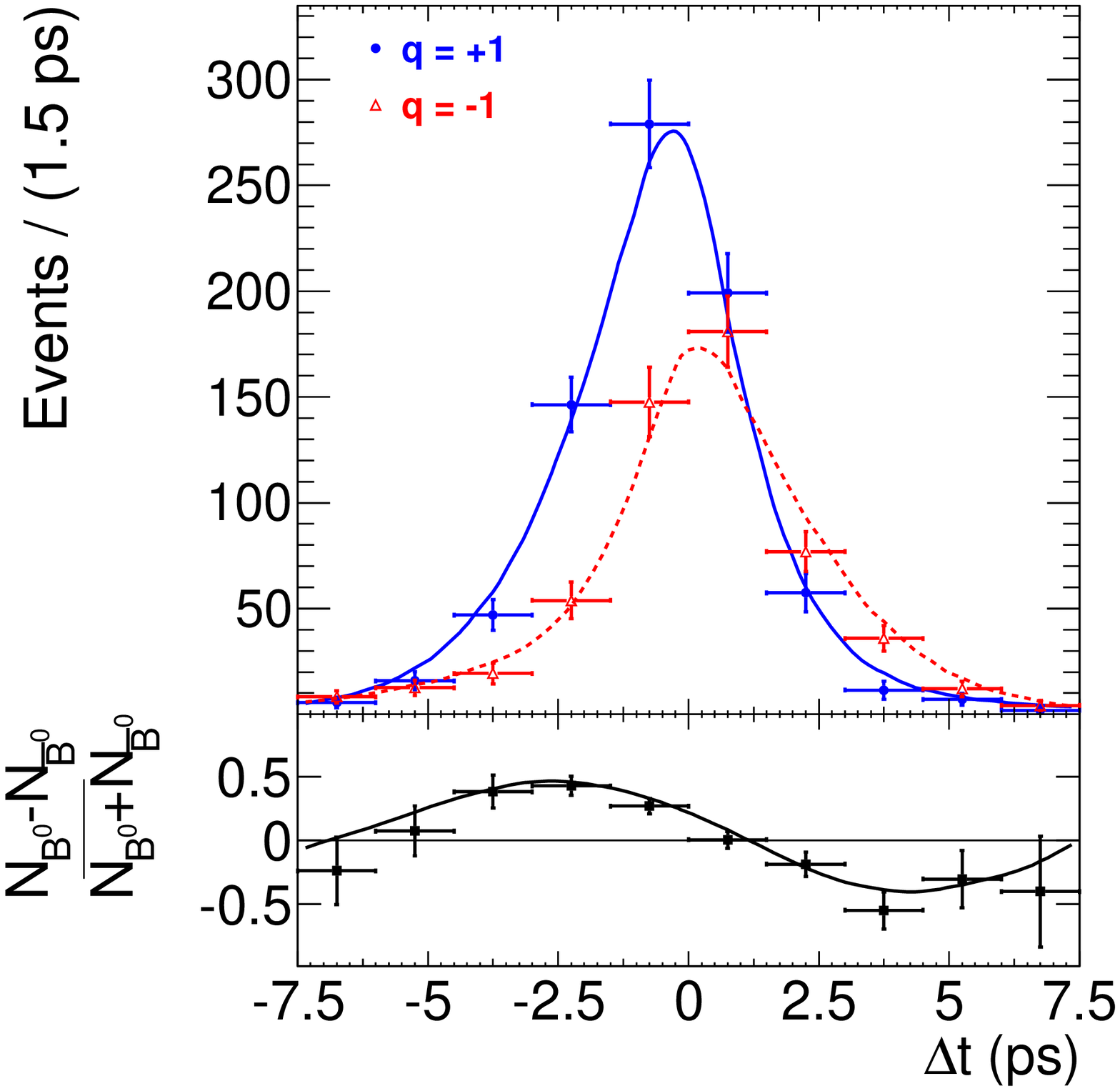}
  \includegraphics[width=.49\textwidth]{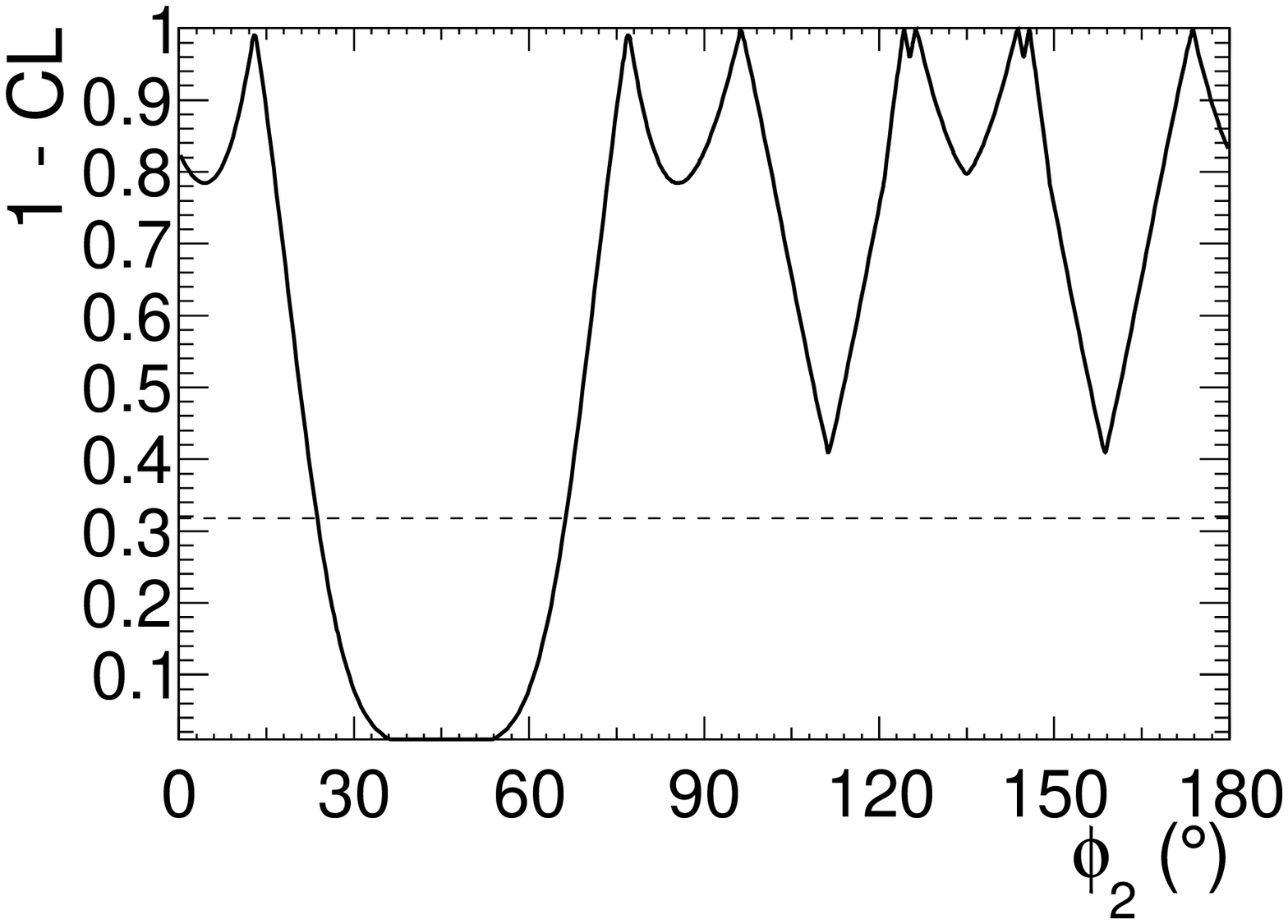}
 \put(-410,20){\scriptsize a)}
 \put(-210,20){\scriptsize b)}\\
  \caption{a) \Dt\ distribution for each flavour tag and the fit result on top and the resulting $CP$ asymmetry for \pippim. Mixing-induced $CP$ violation can be clearly seen in the asymmetry plots and the height difference in the \Dt\ projection indicates direct $CP$ violation. b) scan of \phitwo\ from an isospin analysis in the $B\to \pi\pi$ system, the dashed line corresponds to the one $\sigma$ level.}
  \label{fig_pipi_tcpv}
\end{figure}

Preliminary results of the $CP$ parameters in this pseudo-scalar to scalar scalar decay yield ${\cal S}_{CP} = -0.636 \pm 0.082\;(\rm stat) \pm 0.027\;(\rm syst)$ and ${\cal A}_{CP} = 0.328 \pm 0.061\;(\rm stat) \pm 0.027\;(\rm syst)$. The  $\Delta t$ distributions and the resulting $CP$ asymmetry are shown in Fig.~\ref{fig_pipi_tcpv}~a). Belle excludes the range $\phi_2 \notin [23.8^{\circ}, 66.8^{\circ}]$ at the $1 \sigma$ level by performing an isospin analysis to remove the penguin contribution, see Fig.~\ref{fig_pipi_tcpv}~b). The amount of direct $CP$ violation was found to be smaller compared to the previous measurement at Belle~\cite{pipi_Belle}. The previous result was confirmed with the previous data set of $535\times 10^{6}$ $B\bar{B}$ pairs. The updated $CP$ asymmetries are in better agreement with other experiments~\cite{pipi_BaBar}.

\section{The Decay \rhozrhoz}
\begin{figure}[htb]
  \centering
  \includegraphics[width=.32\textwidth]{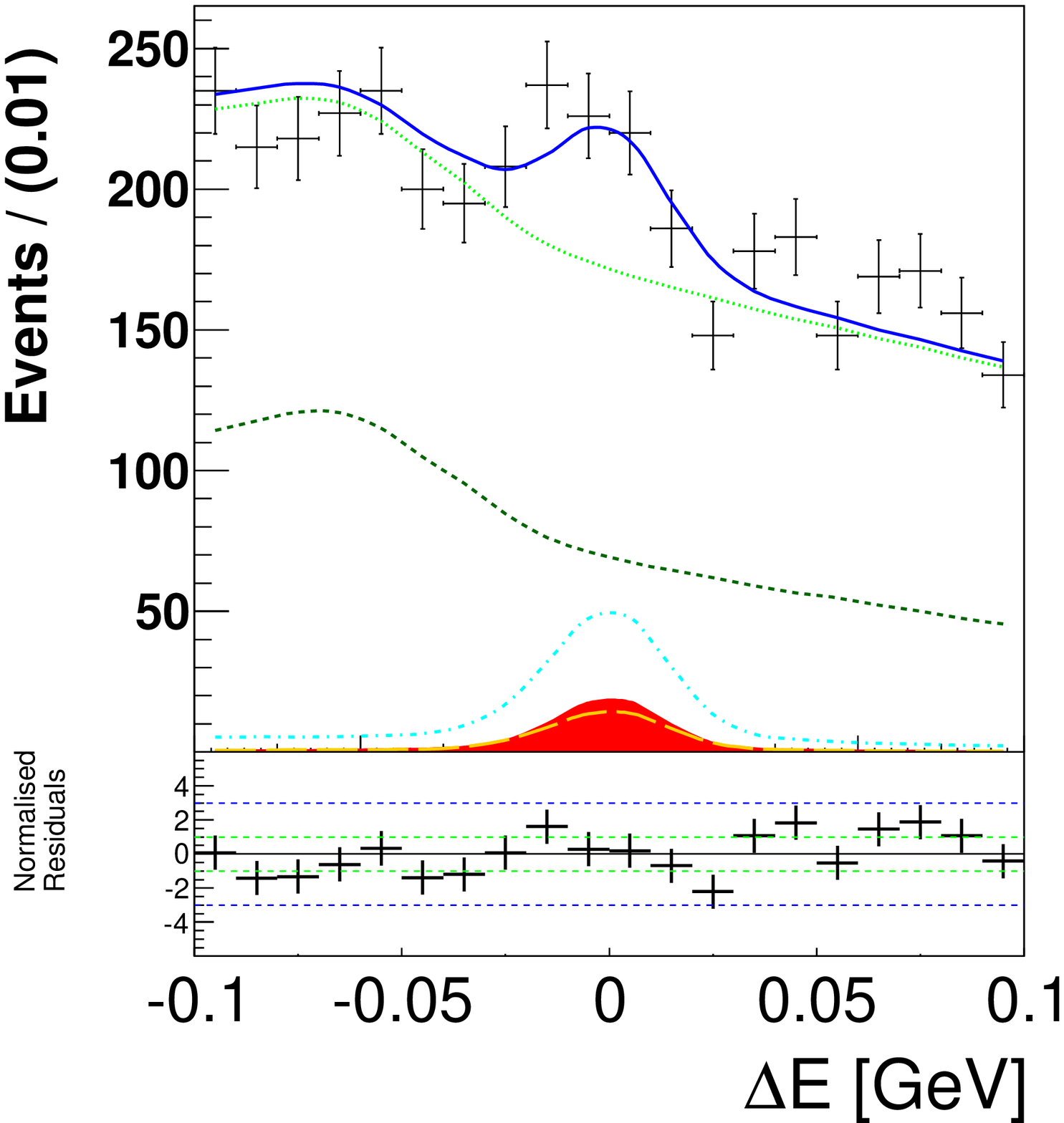}
  \includegraphics[width=.32\textwidth]{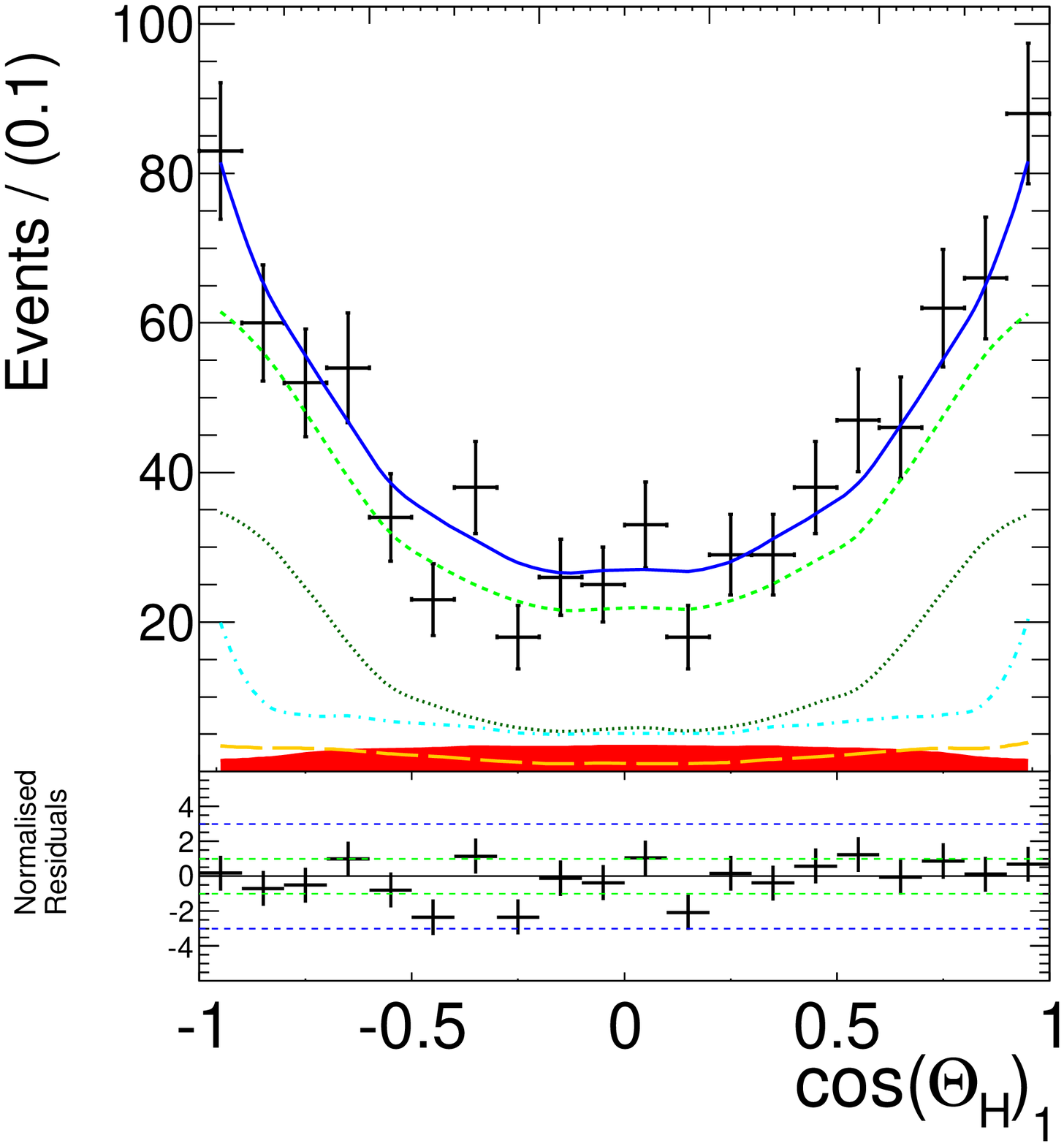}
  \includegraphics[width=.32\textwidth]{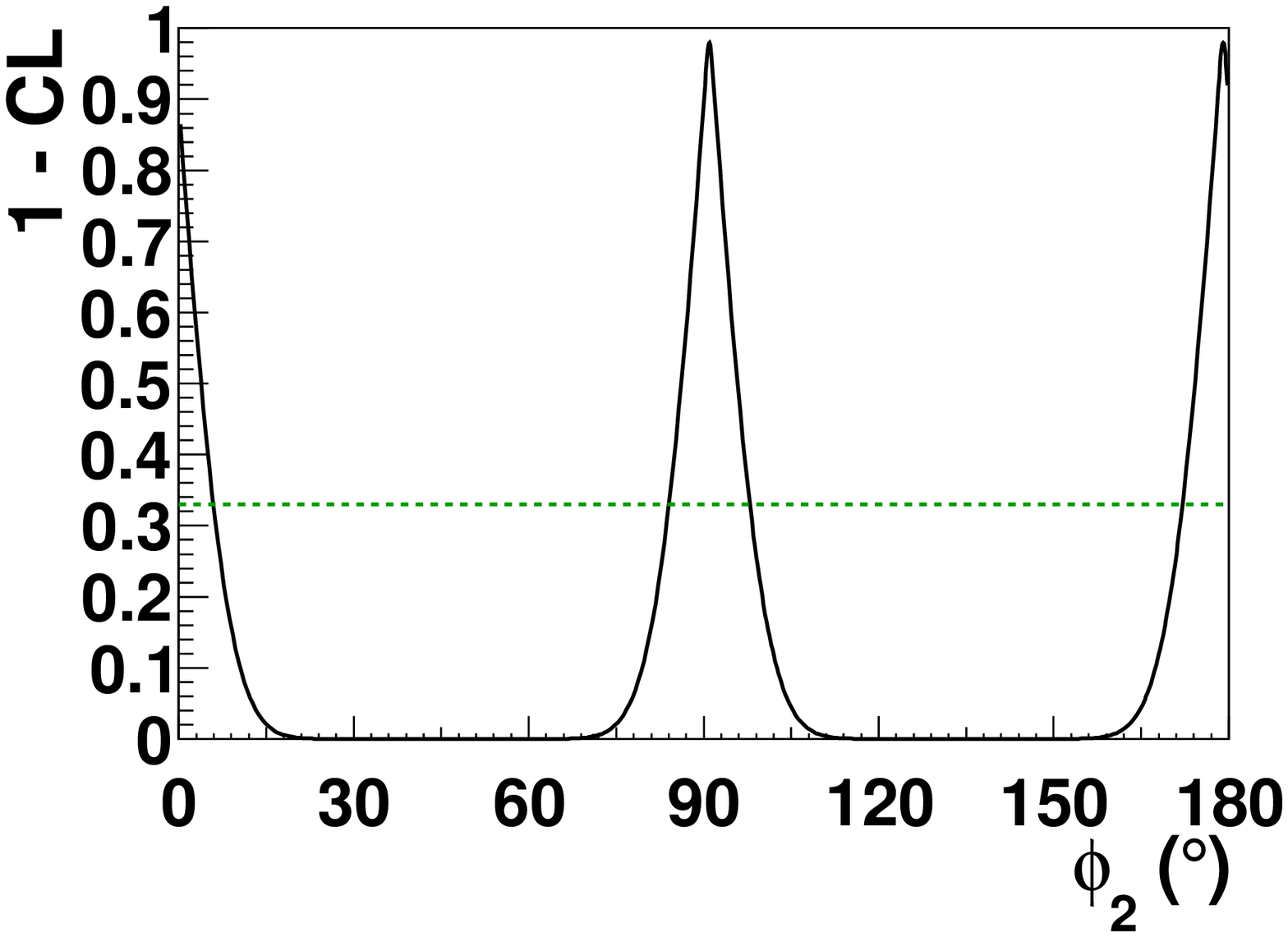}
\put(-410,1){\scriptsize a)}
 \put(-280,1){\scriptsize b)}
 \put(-135,1){\scriptsize c)}\\
  \caption{ Distributions of a) $\Delta E$ and b) $\cos\Theta_{H}$ with the fit result on top. The shaded red area and the long dashed orange histogram are the $B^0\to\rho^0\rho^0$ and $f_0\rho^0$ contributions, respectively. Furthermore, all four pion final states are shown in dashed cyan, the entire ($B\bar{B}$) background in dashed green (dark green) and the full PDF in blue. c) scan of \phitwo\ from an isospin analysis in the $B\to \rho\rho$ system, the dashed line corresponds to the one $\sigma$ level.}
  \label{p_r0r0}
\end{figure}

The presence of multiple, largely unknown backgrounds with the same four charged pions final state as $B^0\to\rho^0\rho^0$ make this rare decay quite difficult to isolate. Interference between the various $4\pi$ modes need to be considered. Having a decay into two vector particles, an angular analysis has to be performed, similar to the decay $B^0 \to D^{*+}D^{*-}$. As for $B^{0}\to\rho^{+}\rho^{-}$~\cite{rhoprhom_BaBar, rhoprhom_Belle1, rhoprhom_Belle2}, the decay \rhorho\ is naively expected to be polarized dominantly longitudinally. However, color-suppressed $B$ decays into light vectors are especially difficult to predict~\cite{BVV_theo}.
Besides updating to the full data set, a helicity angle $\cos\Theta_H$ for each $\rho^{0}$ is added to the fit. The angles, defined in the helicity basis, are powerful in separating the different backgrounds and allow one to measure the fraction $f_L$ of longitudinal (purely $CP$-even) polarization in \rhorho\ decays. As a preliminary result, Belle obtains ${\cal B}(B^0\to\rho^0\rho^0) = (1.02 \pm0.30\;(\rm stat) \pm 0.22\;(\rm syst)) \times 10^{-6}$ with a fraction of longitudinal polarization, $f_L = 0.21^{+0.18}_{-0.22}\;(\rm stat)\pm 0.11\;(\rm syst)$. Having a significance of $2.9$ standard deviations, an upper limit of ${\cal B}(B^0\to\rho^0\rho^0)<1.5 \times 10^{-6}$ at the $90\%$ CL is provided~\cite{belle_r0r0}. Since this mode is currently statistically (and systematically) limited and is found to decay dominantly into transversally polarized $\rho^0$s ($CP$-even and odd), a measurement of the $CP$ asymmetries has not been performed. However, the size of the amplitude of the decays into longitudinally polarized $\rho^0$s from this measurement has been used in an isospin analysis together with world averages of $B^0\to\rho^{+}\rho^{-}$ and $B^+\to\rho^{+}\rho^{0}$ decays~\cite{hfag} (longitudinal polarization only). The resulting constraint consistent with the SM value is $\phitwo = (91.0 \pm 7.2)^{\circ}$. The relatively small amplitude of $B^0\to\rho^0\rho^0$ makes the isospin analysis in the $B\to\rho\rho$ less ambiguous. In addition, Belle reported the first evidence of $B^0\to f_0\rho^0$ decays with a sigificance of $3.0\; \sigma$; ${\cal B}(B^0\to f_0\rho^0) = (0.86 \pm 0.27\;(\rm stat)\pm 0.15\;(\rm syst))\times 10^{-6}$. Distributions of the difference of energy of the reconstructed signal $B$ to the beam energy, $\Delta E$ and one of the helicty angles, each with the fit result on top, are shown in Fig.~\ref{p_r0r0} together with the \phitwo\ scan from the isospin analysis.

Comparing these results with the ones obtained by BaBar, we find good agreement in the branching fraction of $\rhozrhoz$ decays, while there is a $2.1\sigma$ discrepancy in the fraction of longitudinal polarization; BaBar finds $f_L = 0.75^{+0.11}_{-0.14}\;(\rm stat)\pm 0.04\;(\rm syst)$~\cite{r0r0Babar}. Also the branching fraction of $B^0\to f_0\rho^0$ decays is significantly higher then the upper limit provided by BabBar; ${\cal B}(B^0\to f_0\rho^0)<0.34 \times 10^{-6}$. Thus, further studies at higher statistics would be very interesting and hopefully will solve these tensions. 

\section{The Decay \aonepi}
\aonepi\ is another decay with a four charged pion final state sensitive to \phitwo, but is similiar to $B^0\to D^{*\pm}D^{\pm}$, a decay into a non-$CP$ eigenstate. Belle reported the first evidence of mixing induced $CP$ violation in this mode with $3.1\sigma$; $S_{CP} = -0.51 \pm 0.14 \;(\rm stat) \pm \;(\rm 0.08)$~\cite{a1pi_Belle}. The amount of penguin pollution can in general be estimated by using $SU(3)$ symmetry~\cite{CKMangles} but would need more input data. Therefore a scan of an effective angle $\phi_{2}^{eff}$ has been presented, giving a fourfold solution for $\phi_2^{eff} \in [-25.5^\circ,-9.1^\circ], [34.7^\circ, 55.3^\circ]$ and $[99.1^\circ, 115.5^\circ]$, where the $2^{\rm nd}$ interval contains two overlapping solutions. The scan (c) is shown together with the $\Delta t$ distribution (a) and the $CP$ asymmetry (b) in Fig.~\ref{p_a1pi}:

\begin{figure}[htb]
  \centering
  \includegraphics[width=.32\textwidth]{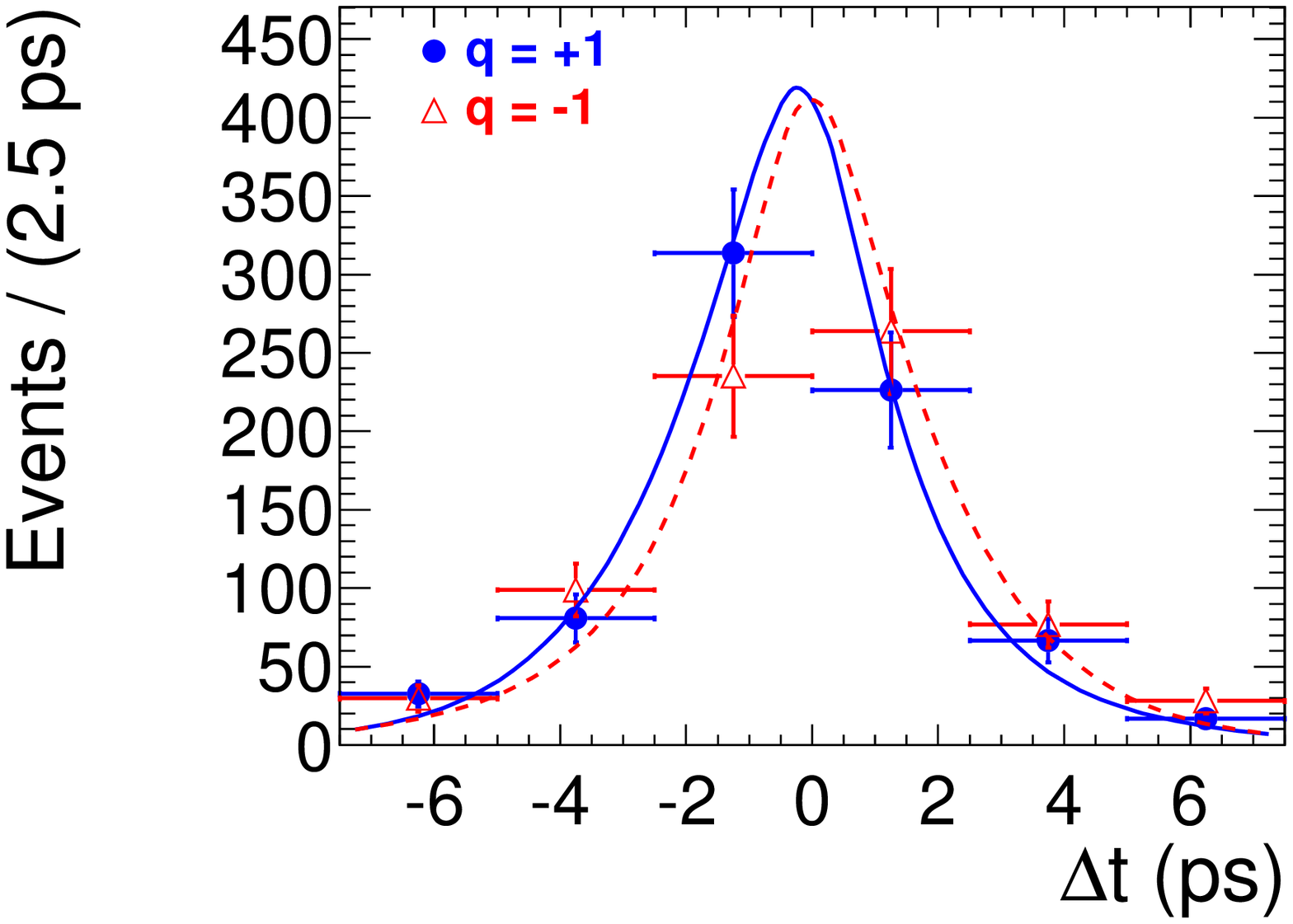}
  \includegraphics[width=.32\textwidth]{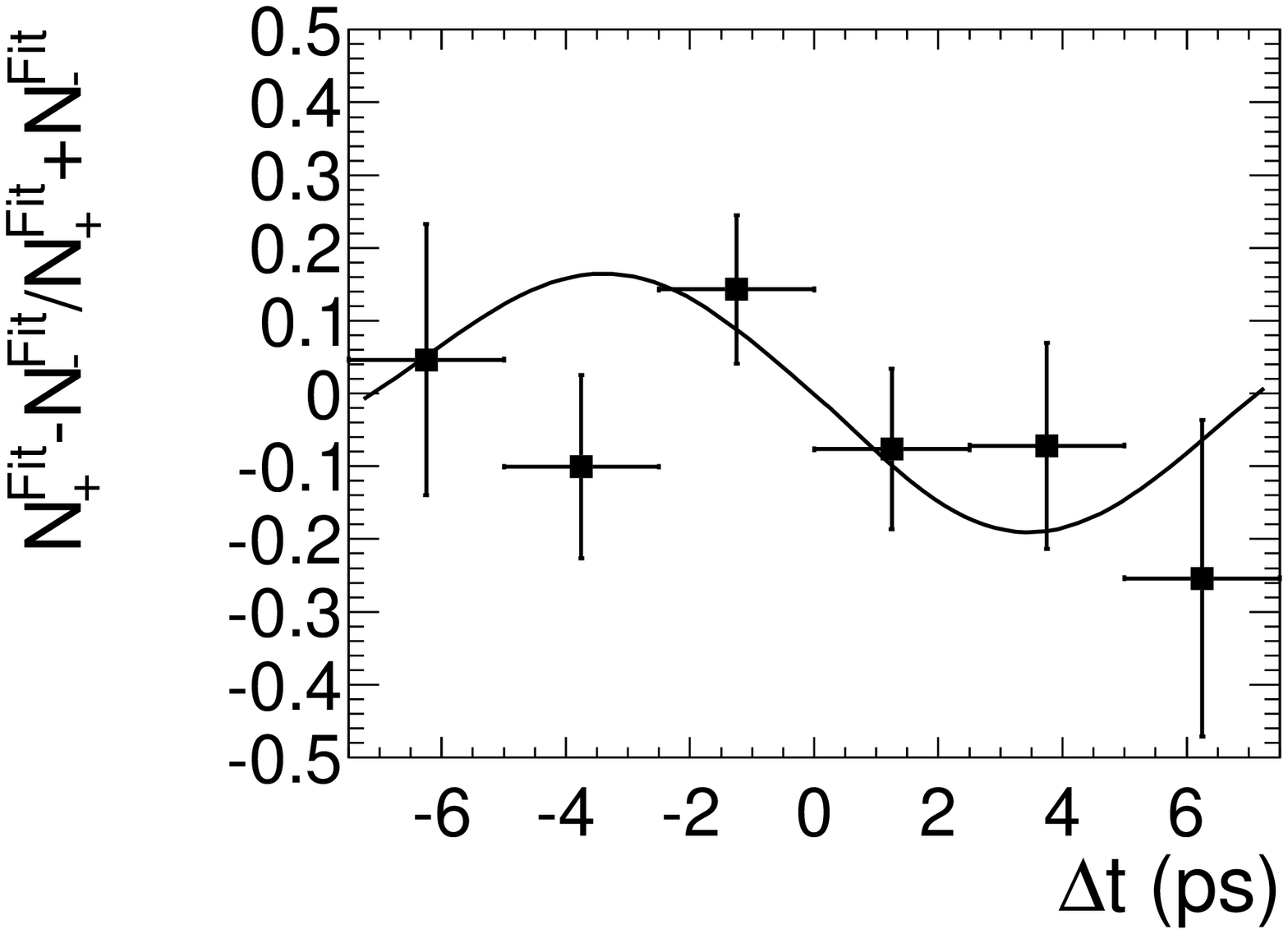}
  \includegraphics[width=.32\textwidth]{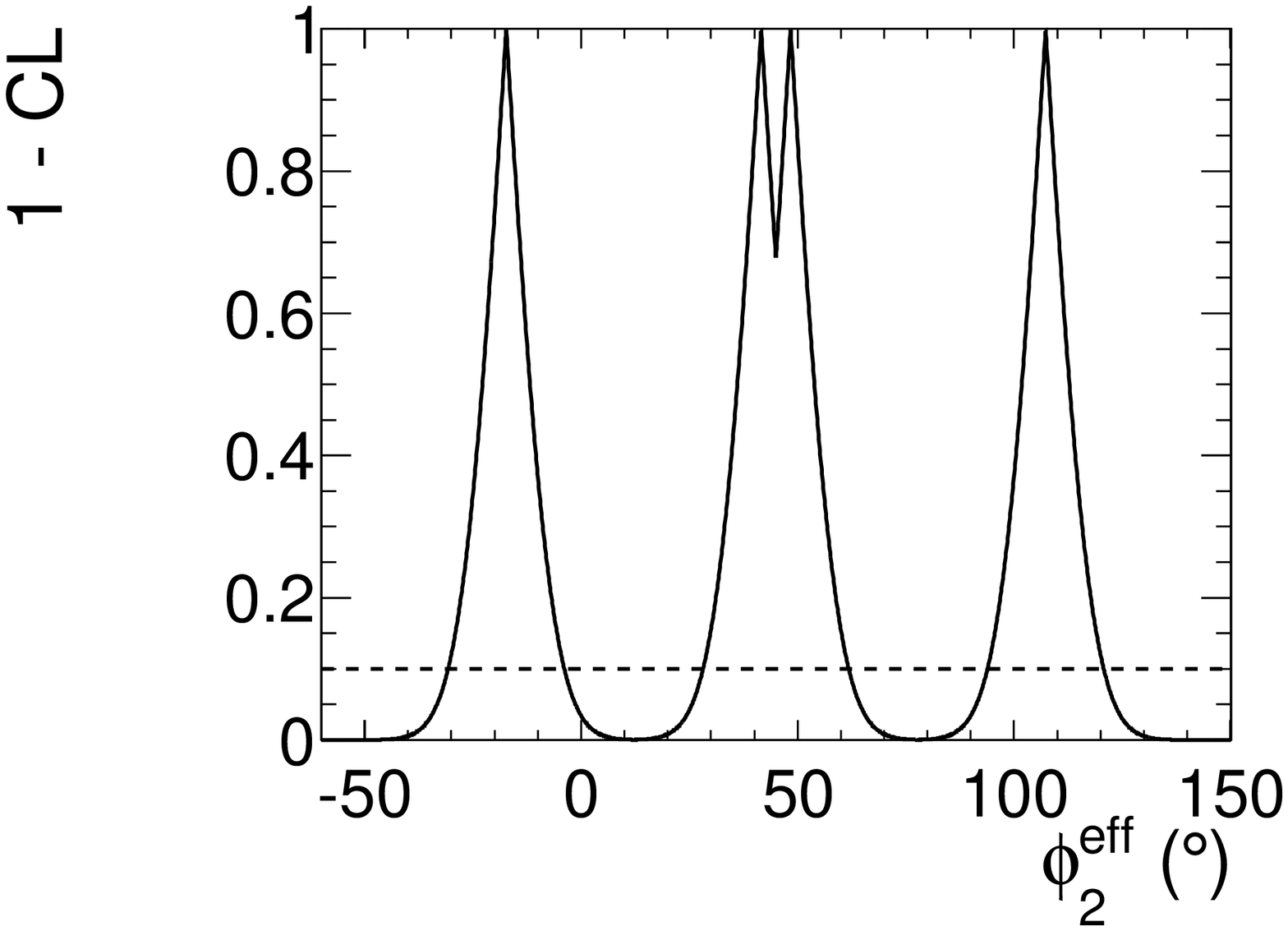}
\put(-400,1){\scriptsize a)}
 \put(-250,1){\scriptsize b)}
 \put(-100,1){\scriptsize c)}\\

  \caption{ a) projection of the fit result onto $\Delta t$ for \aonepi. b) the resulting time-dependent $CP$ asymmetry. c) scan of $\phi_{2}^{eff}$ where the dashed line corresponds to the one $\sigma$ level.}
  \label{p_a1pi}
\end{figure}

\section{Summary}
We have presented recent measurements from Belle sensitive to the CKM phases \phione\ and \phitwo\ using the full data set. For \phione, Belle provides the currently most precise value, $\sin(2\phione) = 0.667 \pm 0.023\;(\rm stat) \pm 0.012\;(\rm syst)$, coming from $B^0\to (c\bar{c})K^0$ decays. Furthermore Belle reported on $CP$ violation in $B\to D^{(*)+}D^{(*)-}$ decays, where mixing induced $CP$ violation in $B\to D^{*+}D^{*-}$ decays was observed for the first time. 

Moreover, preliminary measurements of the $CP$ asymmetries in $B \rightarrow \pip \pim$ and the branching fraction of $B\to\rho^0\rho^0$, together with fraction of longitudinal polarized $\rho^0$s in this decay were presented. The data are used to constrain \phitwo\ with an $SU(2)$ isospin analysis. Also, first evidence of mixing induced $CP$ violation in the decay \aonepi, together with a scan of an effective \phitwo\ was presented. The current world averages of \phione\ and \phitwo\ as computed by the CKMfitter~\cite{CKMfitter} (including the results presented) and UTfit~\cite{UTfit} collaborations are $\phione = (21.73^{+0.78}_{-0.74})^{\circ}$ and $\phitwo = (88.5^{+4.7}_{-4.4})^{\circ}$ and $\phione = (22.28 \pm 0.92)^{\circ}$ and $\phitwo = (89.1 \pm 3.0 )^{\circ}$, respectively. With BelleII being built and LHCb operating, the next generation of $B$ physics experiments are expected to further reduce the uncertainty of the CKM observables, e.g. the uncertainty of \phitwo\ is expected to be reduced to $1^{\circ}-2^{\circ}$~\cite{Belle2}.

\end{document}